\documentclass{elsart3p}

 \usepackage{graphicx}

\usepackage{amssymb}

\begin{document}

\begin{frontmatter}



\title{Role of pair-breaking and phase fluctuations in $c$-axis tunneling in underdoped Bi$_{2}$Sr$_{2}$CaCu$_{2}$O$_{8+\delta}$ }


\author[a]{C.J. van der Beek,}
\author[a]{P. Spathis,}
\author[a]{S. Colson,}
\author[b]{P. Gierlowski,}
\author[c]{M. Gaifullin,}
\author[d]{Yuji Matsuda,}
\author[e]{P.H. Kes}

\address[a]{Laboratoire des Solides Irradi\'{e}s, Ecole Polytechnique, 91128 Palaiseau, France}
\address[b]{Institute of Physics of the Polish Academy of Sciences, 32/46 Aleja Lotnikow, 02-668 Warsaw, Poland}
\address[c]{National Institute for Material Science, 1-2-1, Sengen, Ibaraki, Tsukuba, Japan}
\address[d]{Department of Physics, Kyoto University, Sakyo-ku, Kyoto 606-8502, Japan }
\address[e]{Kamerlingh Onnes Laboratorium, Rijksuniversiteit Leiden, P.O. Box 9506, 2300 RA Leiden, The Netherlands }

\begin{abstract}
    The Josephson Plasma Resonance is used to study the $c$-axis supercurrent 
    in the superconducting state of underdoped 
    Bi$_{2}$Sr$_{2}$CaCu$_{2}$O$_{8+\delta}$  with varying degrees of controlled point-like disorder, 
    introduced by high-energy electron irradiation. As disorder is increased, the Josephson Plasma frequency 
    decreases proportionally to the critical temperature. The temperature dependence of the plasma frequency
    does not depend on the irradiation dose, and is in quantitative agreement with a model for quantum fluctuations
    of the superconducting phase in the CuO$_{2}$ layers.  

\end{abstract}

\end{frontmatter}

\section{Introduction}
\label{Introduction}
From the $d$-wave symmetry of the order parameter of cuprate 
superconductors, one expects an enhanced sensitivity of $c$-axis transport in the superconducting 
state to disorder, due to the enhancement of the quasiparticle density of states along 
the gap node directions, and due to impurity assisted hopping 
\cite{XiangWheatley}. Both  mechanisms lead to as yet unobserved 
$T^{2}$-dependences of the $c$-axis superfluid density $\rho_{s}^c$ 
at low $T$, with coefficients that strongly depend on the scattering rate $\Gamma$.  
On the other hand,  underdoped cuprates are 
sufficiently disordered for (quantum) fluctuations of the  order 
parameter phase to play a prominent role \cite{IoffeMillis}, leading to a $T$-linear 
behavior of $\rho_{s}^{c}$. Here, we report on  
the effect of controlled point disorder on interlayer tunneling of 
Cooper pairs in underdoped Bi$_{2}$Sr$_{2}$CaCu$_{2}$O$_{8+\delta}$.

\section{Experimental details}
\label{Experimental}

Single-crystalline rods of underdoped \linebreak Bi$_{2}$Sr$_{2}$CaCu$_{2}$O$_{8+\delta}$ 
were grown using the travelling solvent floating zone technique under 
25 mBar oxygen partial pressure \cite{MingLi02}. The samples used for the study, 
with $T_{c} \approx 70$ K, were cleaved from the same crystalline piece. 
The crystals were then irradiated with 2.5 MeV electrons using the van 
der Graaf accelerator of the Laboratoire des Solides Irradi\'{e}s. 
The irradiation produces homogeneously distributed 
Frenkel pairs, which have previously been identified as strong 
scattering centers \cite{Florence}.


The Josephson Plasma Resonance (JPR) frequency $f_{JPR}$ of the crystals was then 
measured using the cavity perturbation technique, exploiting $TM_{01n}$ 
harmonic modes to access different frequencies \cite{Colson2003}, and the bolometric 
technique using a waveguide in the $TE_{01}$ travelling wave mode 
\cite{Gaifullin99}. The 
latter technique allows for swept-frequency measurements, necessary to 
elucidate the weak $f_{JPR}(T)$ dependence    at 
low $T$ \cite{Gaifullin99}. Note that $f_{JPR}^{2}$ 
is proportional to the $c$-axis critical current $j_{c}^{c}$ and to the $c$-axis 
superfluid density: $f_{JPR}^{2} =  j_{c}^{c} s/ 2 \pi \epsilon_{0} 
\epsilon_{r} \Phi_{0} = c^{2} / 4 \pi^{2}
 \epsilon_{r} \lambda_{c}^{2} \propto \rho_{s}^{c}$, with $\epsilon_{r}$  the 
 low-frequency dielectric constant and $\lambda_{c}$ is the $c$-axis 
 penetration depth.

\section{Results and Discussion}
\label{Results}

Sharp JPR resonant peaks were measured for all samples under study. 
Both $T_{c}$ and $f_{JPR}(T \rightarrow 0)$ decrease 
with irradiation dose. The sensitivity of $f_{JPR}$ to even weak additional 
disorder contradicts the model of coherent interlayer Cooper pair 
tunneling \cite{Latyshev99}.  Within the framework of a $d$-wave BCS 
model, the measured proportionality between $f_{JPR}^{2}$ and $T_{c}$ can be
understood as resulting either from (i) the dependence of the interlayer Josephson current 
$j_{c}^{c} \propto \Delta \sigma_{qp}$  on the gap magnitude $\Delta$ and the 
quasiparticle conductivity $\sigma_{qp}$ \cite{Sun}, or (ii) from the decrease 
of the \em in-plane \rm superfluid density due to strong phase fluctuations.

\begin{figure}
    \includegraphics[width=7.5cm,keepaspectratio]{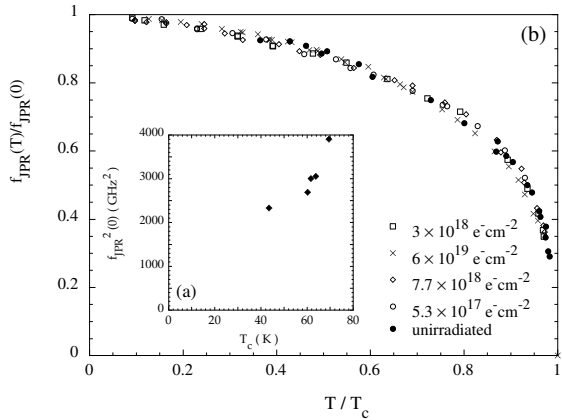}
\caption{JPR frequency of the $e^{-}$-irradiated
Bi$_{2}$Sr$_{2}$CaCu$_{2}$O$_{8+\delta}$ crystals, normalized to the 
JPR frequency at $T \rightarrow 0$, versus reduced temperature.  }
\label{fig:JPR-vs-T}
\end{figure}

The origin of the temperature and disorder-dependence of $f_{JPR}$ can 
be pinpointed using Fig.~\ref{fig:JPR-vs-T}(b). Normalising all results to  
$f_{JPR}(T \rightarrow 0)$ and plotting these versus reduced 
temperature, reveals a common $T$-dependence independent of 
disorder. This contradicts the prediction of 
Ref.~\cite{XiangWheatley} that $\rho_{s}$ should follow different powers 
of $T$ depending on the ratio  of $T$, $\Delta$, and $\Gamma$. 
However, it agrees with a dominant role of quantum phase fluctuations. Then, 
\begin{equation}
\frac{f_{JPR}^{2}(T)}{f_{JPR}^{2}(0)} \approx 1 - \frac{4 \alpha
k_{B}T}{\varepsilon_{0}s \overline{\sigma} } - C_{2} 
\frac{2 \overline{\sigma} }{3} \left(\frac{k_{B}T}{\varepsilon_{0}s 
}\right)^{2}
\label{eq:paramekanti}
\end{equation}
\noindent where $\alpha = (\partial \varepsilon_{0}s / \partial 
T)_{T\rightarrow 0}$, $\varepsilon_{0} = 
\Phi_{0}/4\pi\mu_{0}\lambda_{ab}^{2}$ with $\lambda_{ab}$ the 
in-plane penetration depth, $s$ is the CuO$_{2}$ layer spacing, and
$\overline{\sigma} = \sigma s$ is the CuO$_{2}$ plane sheet 
conductivity \cite{Paramekanti}. The temperature dependence of the 
JPR frequency arises from the temperature of the in-plane phase 
stiffness $\varepsilon_{0}$. Figure~2 shows that 
Eq.~(\ref{eq:paramekanti}) describes the results very
satisfactorily.

Summarizing, the temperature- and disorder dependence of the $c$-axis Cooper 
pair tunnel current in underdoped Bi$_{2}$Sr$_{2}$CaCu$_{2}$O$_{8+\delta}$ 
is well described assuming a strong effect of quantum phase 
fluctuations. A $d$-wave model without fluctuations cannot account for 
the $T$-dependence of $c$-axis coupling.

\begin{figure}
    \includegraphics[width=8.0cm,keepaspectratio]{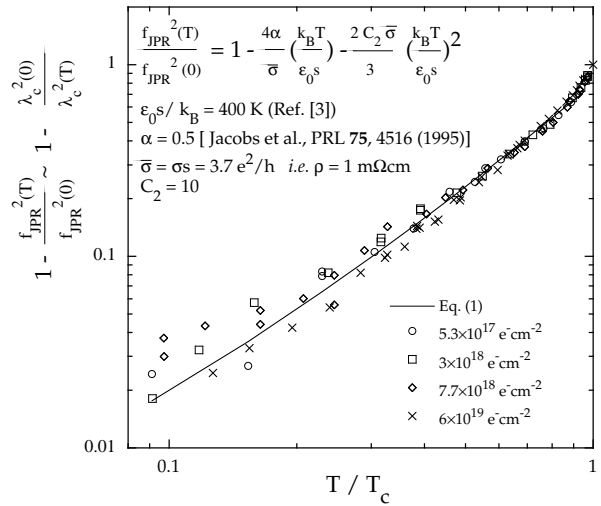}
    \label{fig:normalize}
\vspace{-4mm}
\caption{JPR frequency of the $e^{-}$-irradiated
Bi$_{2}$Sr$_{2}$CaCu$_{2}$O$_{8+\delta}$ crystals, plotted as $1 - 
f_{JPR}^{2}(T)/f_{JPR}^{2}(0)$ in order to bring out the low 
temperature $T$-dependence. The line is a fit to Eq.~\protect\ref{eq:paramekanti} 
\protect\cite{Paramekanti} with 
parameters as indicated.}
\end{figure}

%
%
%

\end{document}